\documentclass{pnastwo}

\usepackage{amssymb,amsfonts,amsmath}

\usepackage{graphicx}
\usepackage{dcolumn}
\usepackage{bm}

\contributor{Submitted to Proceedings
of the National Academy of Sciences of the United States of America}
\url{www.pnas.org/cgi/doi/10.1073/pnas.0709640104}
\copyrightyear{2008}
\issuedate{Issue Date}
\volume{Volume}
\issuenumber{Issue Number}

\begin{document}


\title{Heat transfer from nanoparticles: a corresponding state analysis}

\author{Samy Merabia
\affil{1}{Universit\'e de Lyon; Univ. Lyon I,  Laboratoire de
Physique de la Mati\`ere Condens\'ee et des Nanostructures; CNRS,
UMR 5586, 43 Bvd. du 11 Nov. 1918, 69622 Villeurbanne Cedex,
France},
Sergei Shenogin\affil{2}{Rensselaer Nanotechnology Center and Department of Materials Science and Engineering,
Rensselaer Polytechnic Institute, Troy, New York 12180, USA},
Laurent Joly \affil{1},
Pawel  Keblinski\affil{2},
\and Jean-Louis Barrat\affil{1}{}}

\contributor{Submitted to Proceedings of the National Academy of Sciences
of the United States of America}

\maketitle

\begin{article}

\begin{abstract}
 In this contribution, we study situations in which nanoparticles in a fluid are strongly
 heated, generating high heat fluxes. This situation is relevant
to  experiments  in which a fluid is locally heated using selective absorption of radiation by solid particles. We first study this situation
for  different types of molecular interactions, using models for gold particles suspended in octane and in water. As
already reported in experiments, very high heat fluxes and temperature elevations (leading eventually to particle destruction)
can be observed in such situations.  We show that a very simple modeling based on Lennard-Jones interactions captures
the essential features of such experiments, and that the results for various liquids can be mapped
onto the Lennard-Jones case,  provided a physically justified (corresponding state)  choice of parameters is made. Physically, the
possibility of sustaining very high heat fluxes
 is related to the strong curvature of the interface that inhibits the formation of an insulating vapor film.
 \end{abstract}


\section{Introduction}
Sub-micron scale heat transfer is attracting a growing interest, motivated by both
 fundamental and technological points of view. In fluids, considerable attention has been
 devoted to the so called nanofluids \cite{eastman2004,prasher2008},
 in which nanoparticles in dilute suspension appear to modify both
  bulk heat transfer and critical heat fluxes. While the former effect can  presumably be understood in terms
  of particle aggregation \cite{prasher2006,vladkov2008}, the latter is still poorly understood.

  More generally, heat transfer from nanoparticles or nanostructures to a fluid
  environment is a subject of active research, stimulated
  by the development of experimental techniques such as time resolved
  optical absorption or reflectivity, or photothermal correlation spectroscopy \cite{radunz2009}.
  Applications  include e.g.  the enhancement of cooling from structured
  surfaces, local heating of fluids by selective absorption from nanoparticles, with possible biomedical hyperthermia
 uses \cite{hirsch2003,hamaguchi2003}.  Recent experiments demonstrated the possibility of reaching very high local temperatures using
 laser heating of nanoparticles \cite{hartland2004,plech2004,kotaidis2006}, even reaching the melting point of gold particles suspended in water.
 From a conceptual point of view, such experiments raise many interesting questions compared to usual, macroscopic
 heat transfer experiments.  How are  the phase diagram and  heat transfer equations modified at small scales ? How
 relevant is the  presence of interfacial resistances and how do they change with temperature ?

 The case of nanofluids\cite{wang2007} is a good illustration of the role that can be played by molecular simulation in the interpretation
 of such complex situations. While many interpretations have been proposed to explain the reported experimental results, it
 is only simulation of simple models that has been able to disprove some of these interpretations and to demonstrate the validity of
 the alternative, aggregation scenario. Interestingly, the use of complex models with accurate interaction force fields is not, in general,
 needed to answer the basic qualitative questions raised by such novel experimental approaches.

 In this manuscript, we use molecular simulation to study the heat transfer from solid nanoparticles to a surrounding fluid
 under extreme conditions (high heat fluxes) using both realistic and simplified molecular models. We show that, in agreement with experiments,
 the temperature of nanometer sized particles can be elevated considerably without inducing bubble nucleation in the fluid.
  This feature is contrasted with the situation for flat surfaces, at which an instability leading to the formation of an insulating
  vapor layer takes place at much smaller  heat fluxes and temperatures \cite{garrison2001,merabia2009}. Using a comparison between a "realistic"
  description of gold-octane and gold-water  systems  and of a simplified Lennard-Jones model, we show, based
  on a  "corresponding state"
  analysis, that the features observed are quite universal. A simple mapping using critical temperature, interfacial
  and heat conductivity properties of the particle/solvent pair allows one to  reproduce accurately the behavior of different systems.

\section{Gold nanoparticles in octane}
\label{gold}
\begin{figure}[htbp]
\begin{center}
\includegraphics[width=6cm]{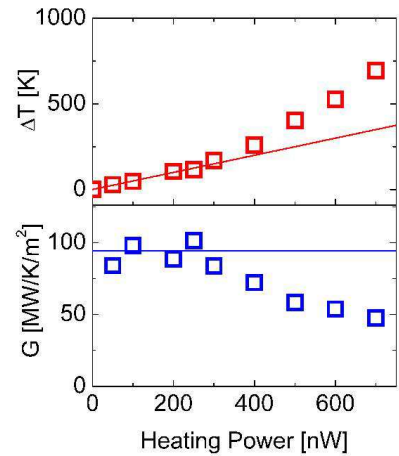}
\caption{\label{conductance_gold} (Color online) Temperature drop at the interface for octane-gold model system (top panel points) and calculated interface conductance (bottom panel points) as a function of heating power. Lines represent linear response regime.
}
\end{center}
\end{figure}

\begin{figure}[htb]
\begin{center}
\includegraphics[width=6cm]{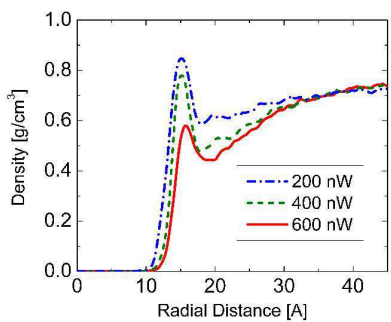}
\caption{\label{density_gold} (Color online) Steady state octane density profiles for octane-gold model system at three heating power levels.}
\end{center}
\end{figure}

To establish a connection with  experimental reality, we began by using molecular
dynamics to simulate heat transfer from a model gold nanoparticle into octane solvent.
The selection of a relatively simple organic solvent in this first part, rather than water, is motivated by
the fact that molecular models are quite accurate in predicting liquid-vapor phase diagrams for alkanes.
Our model consists of a nanoparticle, with an average radius of about 1.3 nm, containing 494
gold atoms arranged on a FCC lattice with density 19.5 g/cm$^3$. The nanoparticle
is immersed in liquid octane containing 2721 octane molecules (21768 united carbon-hydrogen
 atoms) and placed in a cubic simulation cell with periodic boundary conditions.
 The interactions between united atoms forming octane molecules are described by
 the Amber force-field~\cite{cornell1995} with all non-bonded interaction energy calculated
 according to 6-12 Lennard-Jones (LJ). The interaction between gold atoms was also
 described by a 6-12 LJ potential, fitted to reproduce bulk Au density (19.5 g/cm$^3$) and the melting point (1310 K).
 Finally, the interaction potential between octane united atoms and gold atoms was
 adapted from Ref.~\cite{xia1992}. All simulations were carried out at constant
 pressure of 1 atmosphere and with the integration time step of 2 fs.
In the equilibration stage of the simulation, a global thermostat is used to maintain
 the overall temperature at 300 K.  The equilibration stage takes approximately 1 ns,
  and under 1 atm, the equilibrated system is contained in a cubic simulation box
  with the edge length of about 90 $\AA$. The nanoparticle is initially placed in the center
  of the simulation cell, but is allowed to freely diffuse during the simulations.
  . The equilibrium
   density of the model octane fluid  is equal to 0.71 g cm$^{-3}$, which compares
   well with the experimental density of 0.7025 g cm$^{-3}$.
To study the heat flow from the nanoparticle to the solvent, the nanoparticle was heated,
with a constant heating power in the range of 100-1000 nW, by rescaling of the
 atomic velocities every time step. The liquid octane in the periphery of the system,
 at distances more than 40 \AA \ from the nanoparticle center (taking into account a possible diffusive motion),
  was maintained at 300 K,
 thus providing the heat sink.
Up to heating powers of 700 nW, after a transient of about 100 ps, a steady state is established.
In the steady state, we collect time averages (over 1ns) of density and temperature profiles
obtained for spherical shells concentric with the nanoparticle center and with a thickness of
 2\AA. For heating powers larger than 700 nW, the system was unstable
and its behavior will be described below.
\begin{figure}[htb]
\begin{center}
\includegraphics[width=6cm]{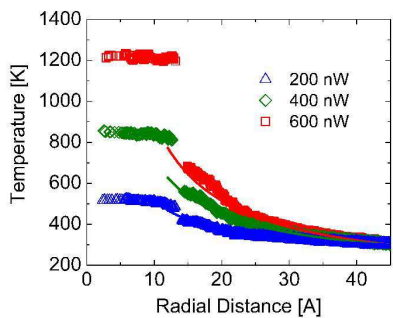}
\caption{\label{temp_gold} (Color online) Steady state temperature profiles for octane-gold model
system at three heating power levels (symbols) and fits with the continuum
theory prediction $T(r)=A+B/r$ (solid lines)}
\end{center}
\end{figure}
Steady state temperature profiles for P = 200, 400 and 600 nW heating powers
are presented in Fig.~\ref{temp_gold}. The temperature profiles have several noteworthy
features. First, the temperature of the nanoparticle is more or less uniform; this is an effect
of the relatively high thermal conductivity of crystalline solid, as compared to liquid. We mention here, that
the electronic contribution to the conductivity,  which provides the dominant mechanism of heat conduction within metallic nanoparticles,
is not accounted for in our description. The essential point, however,  is that the conductivity of the solid is much higher than that of the liquid,
independently of the precise mechanism involved.
Second, the temperature in the liquid follows the solution of the continuum heat flow
 problem with spherical symmetry (see solid lines in Fig.~\ref{temp_gold}). In the steady
 state the temperature profile is described by the solution of the Laplace equation in the form:
\begin{equation}
\label{temp_spherical}
T(r)=A+B/r
\end{equation}
Near the particle-liquid interface, the temperature profile deviates from the formula
given by the Eq.~(\ref{temp_spherical}), particularly for larger heating powers involved.
 This deviation is likely due to the non-uniform thermal transport properties of the liquid,
  since Eq.~\ref{temp_spherical} is valid under the assumption that the thermal conductivity is constant.
Very importantly, there is a large temperature drop, $\Delta T$, at the nanoparticle-liquid interface which is a manifestation of the interfacial thermal resistance. Such resistance is caused by the mismatch of thermal properties between the solid and liquid components, and is also affected by the strength of the interfacial bonding. The interfacial thermal conductance, $G$, can be quantified via the relationship:
\begin{equation}
j_Q=G \Delta T
\end{equation}
Where $j_Q$ is the heat flux across the interface and $\Delta T$ is the discontinuous temperature at the interface (see temperature profiles in Fig.~\ref{temp_spherical}). Fig.~\ref{conductance_gold} (top panel) shows the relationship between the heating power $P$ and the temperature drop at the octane-liquid interface. At lower heating powers the heat flux is proportional to the heating power, indicating constant value of the interfacial conductance. However, above the 300nW heating power, the increase in the temperature drop becomes steeper, indicating increasing interfacial thermal resistance (see Fig.~\ref{conductance_gold}).
The calculated interfacial conductance as a function of $P$ is shown in Fig.~\ref{conductance_gold} (bottom panel). At small heating powers (temperature drops), the value of the interfacial conductance is about 100 MW/$m^2$/K. This value is similar to those obtained in an experiment on gold nanoparticle-water dispersions~\cite{wilson2002}. With increasing heat power (temperature drop), the interfacial thermal conductance decreases from 100 to about 50 MW/m$^2$/K at $P = 700$nW. To gain an insight into the structural origin of the behavior, we show   in Fig.~\ref{density_gold} the octane density profiles corresponding to temperature profiles from Fig.~\ref{temp_gold}. As the temperature of the nanoparticle and the adjacent liquid increases, there is a visible decrease of liquid density adjacent to the
solid surface (see Fig.~\ref{density_gold}).  This increase of the molecular distance between liquid molecules and solid atoms is likely responsible for the decrease of the interfacial thermal conductance.

\section{Gold nanoparticles in water}
\label{gold2}
In order to explore a different type of bonding for the liquid, and also to allow a connection to recent experiments \cite{radunz2009},
we explore in this section a system made of a gold nanoparticle similar to the one described in the previous section, solvated in water modeled
using the standard SPC/E model \cite{vega}. In order to  allow comparisons with the Lennard-Jones calculations described below, we studied water pressurized
 at a pressure of 80 bars, and at a temperature of $450$ K. These parameters were chosen so that the distance
 to the critical point is large, but still within a range that allows comparisons between different models.
  The system was made of 10000 water molecules.

 The main issue here is the choice of the model for the gold-water interaction.
Experimentally, the results reported for the wetting of water on gold are quite scattered,
with in general a large contact angle hysteresis. Results obtained under UHV conditions \cite{bewig1965,schrader1970} report
 a low contact angle (smaller than  30 degrees, or close to a wetting situation).  On the other hand, the force fields that exist in the literature, and have been
 based on density functional theory DFT calculations \cite{garrison2001,kremer2005},   yield  higher contact angles

 In this study we make the choice of strengthening  the attractive terms in this effective potentials,
 in order to obtain a contact angle  (estimated from a simple calculation of the Lennard-Jones contributions to the surface
 tensions \cite{widom1982},  and of a the actual value of the SPC/E surface tension at 300K)
 of   the order of 25 degrees, consistent with experiments. As a result, the gold-water interaction is written in the form
 of a standard 6-12 potential like in \cite{kremer2005}, with the following
 parameters: $\epsilon_{O/Au}=0.59$ kcal.mol$^{-1}$ and $\sigma_{O/Au}=0.36$ nm,
   while the hydrogen atoms do not interact with the gold atoms.
The system NP + water is first equilibrated during 100 000 time steps which represent a
physical time of $200$ ps. Then the nanoparticle is heated up at a constant
power while the water molecules at a distance 20 \AA from the nanoparticle are
thermostated at 450K. In all the following, we will restrict ourselves to moderate
heating powers (smaller than 700 nW), since for larger heating intensities, we
have observed non-stationary effects in the heat transfer process. While these effets are
interesting in themselves, their study is out of the scope of this article, and we
leave a complete study for future work.
\begin{figure}[htb]
\begin{center}
\includegraphics[width=6cm]{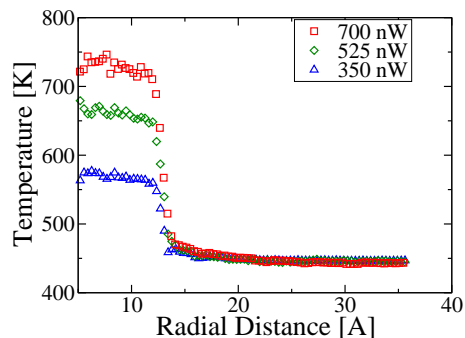}
\caption{\label{temp_gold_water} (Color online) Temperature profiles across the
  water-gold nanoparticle interface at three heating strengths.}
\end{center}
\end{figure}
Temperature profiles of gold NP immersed in water, heated at various powers,
are shown in figure \ref{temp_gold_water}. Clearly, temperature profiles are flatter in water than in
octane, if the comparison is made at the same value of the heating power.
 This is due to the about 5 times  larger conductivity of water compared to octane.
On the other hand, interfacial temperature jumps $\Delta T$ are smaller in the water/gold
case.
Note that it is essential to do the
comparison at a given value of the heating power, not at a given value of the
nanoparticle temperature. For instance, for the
nanoparticle heated up at $400$ nW in octane, we have measured $\Delta T = 220$ K,
while it is a factor of 2 less if
the nanoparticle is immersed in water. Consequently, the water/gold
interface has a larger interfacial conductance than the octane/gold system.
The value we have measured varies from $G=170$ MW/m$^{2}$/K  to $G=150$ MW/m$^{2}$/K
over  the range  of heating power investigated, a variation that is smaller than in the case of octane. This trend
is consistent with the recent finding that the interfacial conductance increases with work of adhesion, which is higher for the gold-water
case than for the gold-octane case. \cite{shenogina2009}
The Kapitsa length
$l_K=\lambda/G$, where $\lambda$ is the thermal conductivity, is of the order of $3.4$nm in this case and of $1$nm in the gold/octane case.

\section{Melting of the nanoparticles}
\label{melting}
\begin{figure}[htb]
\begin{center}
\parbox{10cm}{\includegraphics[width=4cm]{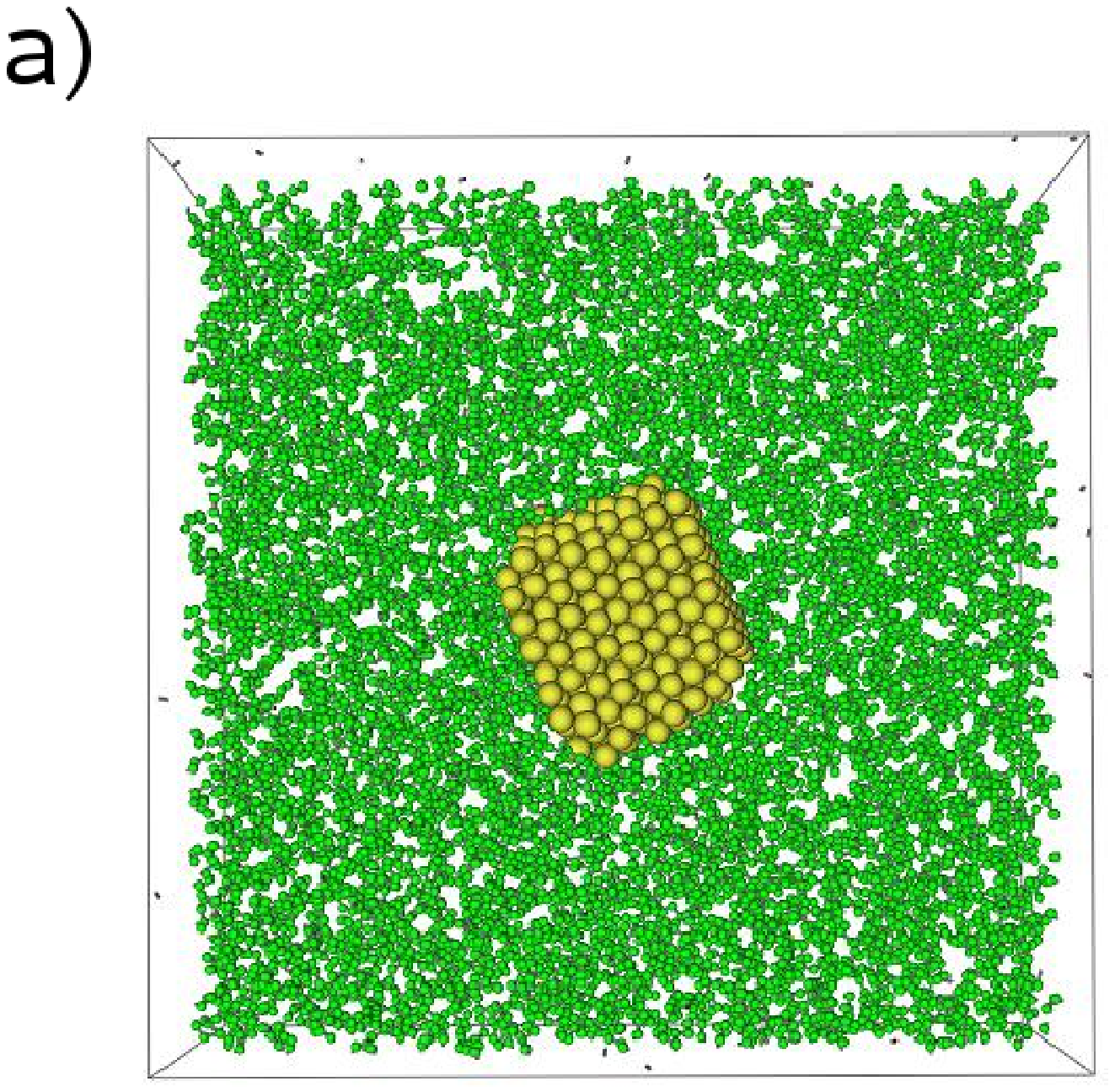}\includegraphics[width=4cm]{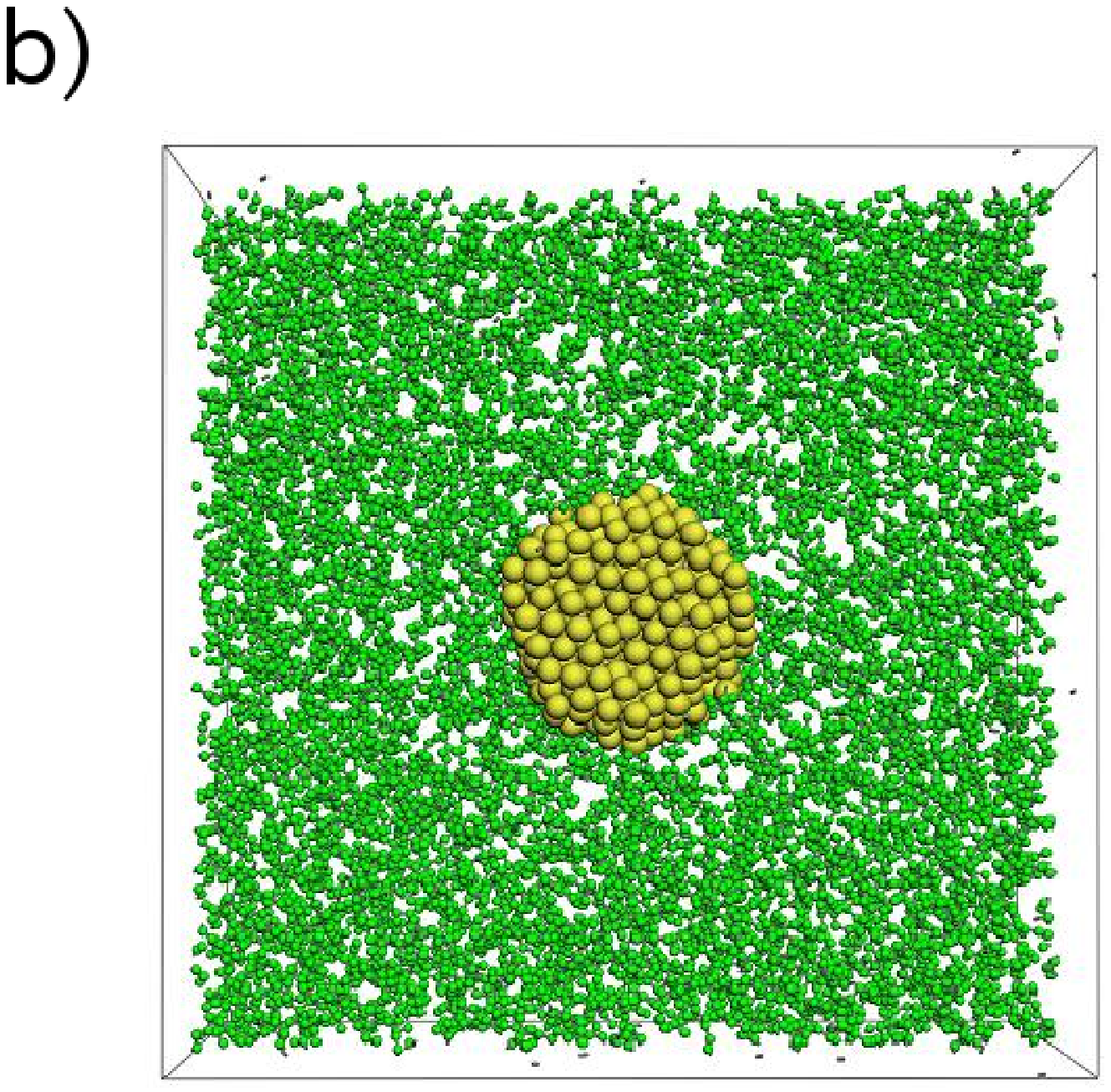}}
\parbox{10cm}{\includegraphics[width=4cm]{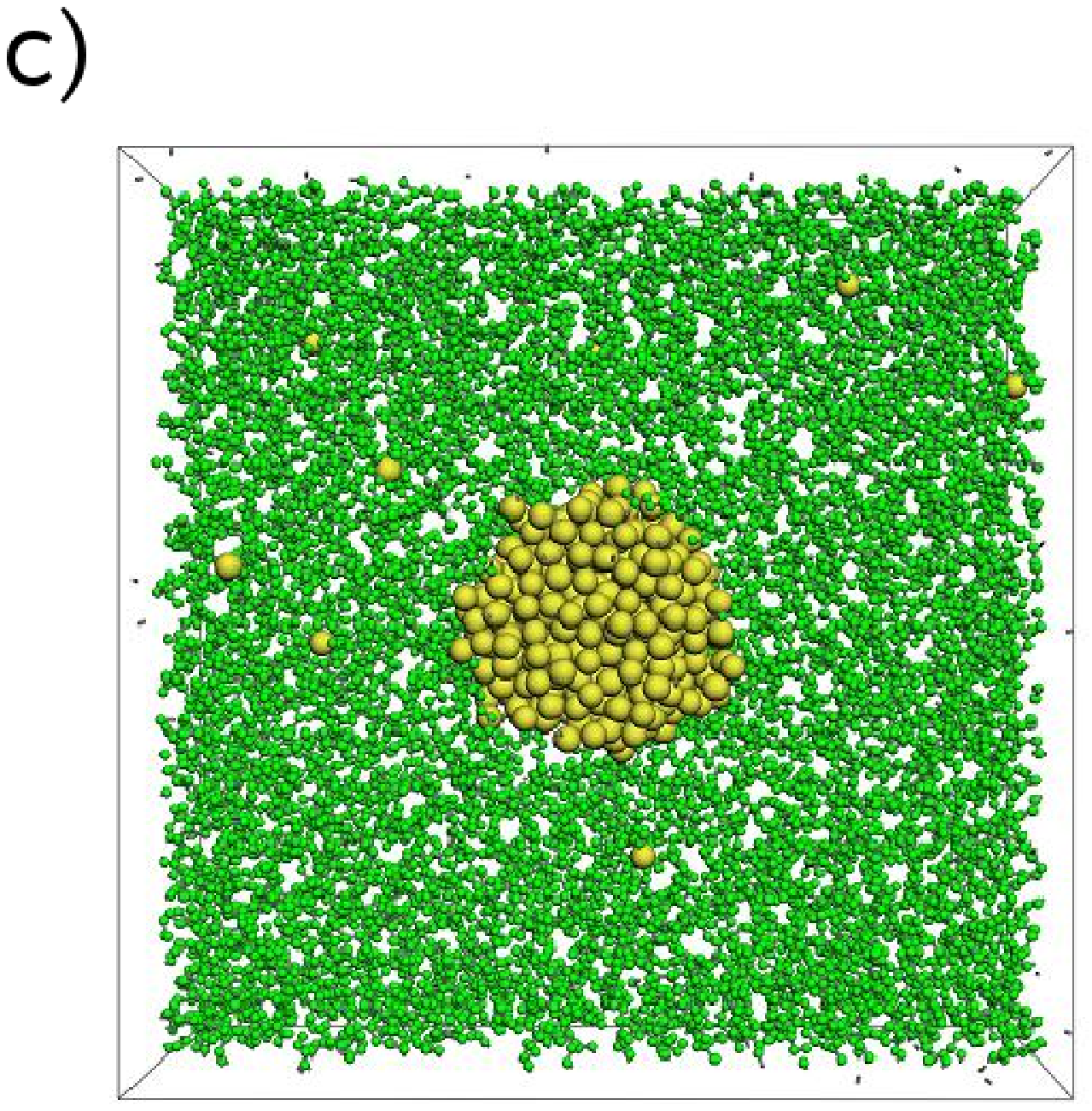}\includegraphics[width=4cm]{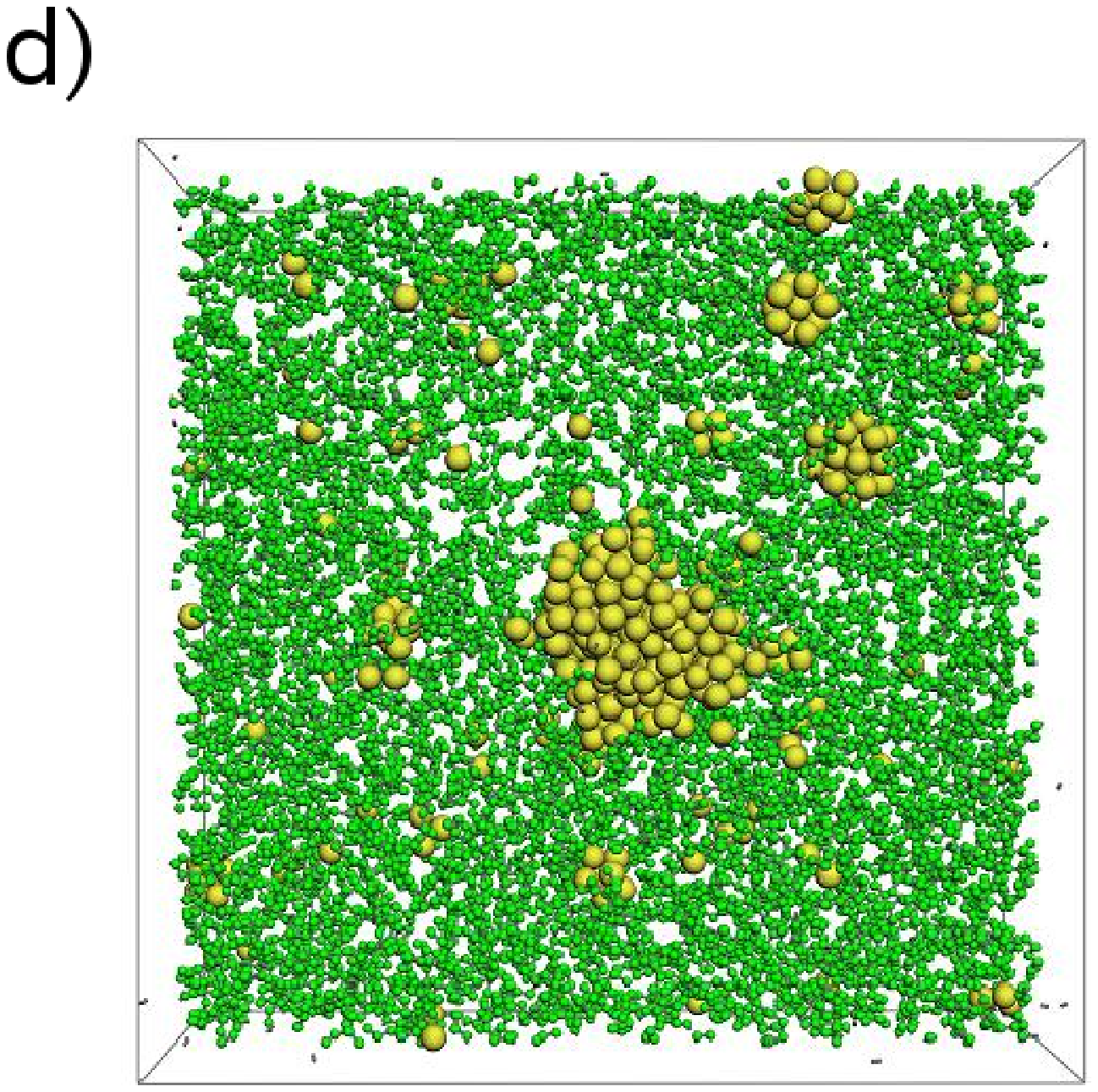}}
\caption{\label{snapshots} (Color online) Snapshots of octane-gold model system at heating powers of 0 nW (a), 500 nW (b), 700 nW (c) and 1000 nW (d).
}
\end{center}
\end{figure}

At high enough temperatures, experiments have  illustrated the possibility of particle melting ~\cite{plech2004} within the fluid. We also explored
briefly this issue in our simulations of gold  particle in octane.
By inspection, we observed that up to P = 500 nW, the nanoparticle structure
remained crystalline. However, at P = 600 nW, the crystalline order of the
nanoparticle is lost. At P = 700 nW, we observed that atoms from the
nanoparticle surface are gradually  "evaporated"  into the solution. At a
later stage they recombine into small Au clusters (Fig.~\ref{snapshots}). We
note that all these processes occur without  formation of a liquid vapor interface, which, as
discussed below, is caused by very large Laplace pressure. In fact, using the
surface tension of octane at room temperature $\gamma=21.8$ $10^{-3}$ N/m and a
bubble radius of $R_0$=2 nm, one obtains the Laplace pressure of
$P_L$=2$\gamma$/$R_0$ $\simeq$ 200 atm. This value is much larger than  the
critical octane pressure of 25.5 atm. These results illustrate an exciting
possibility of decomposition of metal nanoparticles into metal atoms or small
clusters, without explosive evaporation or thermal damage on the embedding
medium. Interestingly, we have not observed such fragmentation for
nanoparticles immersed in water. Although the crystalline order is lost, the
gold nanoparticle keeps its integrity. This is probably due to the higher
interfacial energy of the gold/water interface compared to gold/octane.

\section{Lennard-Jones model}
\label{lj}
To emphasize the generality of the scenario described above, we briefly recall here the results
obtained  for a generic model of a simple, Lennard Jones monoatomic fluid in contact
with a heated solid particle \cite{merabia2009}.
The  system is made  of
a FCC solid particle formed of 555 atoms, immersed in a fluid of 23000 atoms, all the atoms interacting through a Lennard-Jones
potential $V_{\alpha \beta}(r)=4 \epsilon ((\sigma/r)^{12}-c_{\alpha \beta} (\sigma/r)^6)$
   where ${\alpha, \beta}$ refers to solid or liquid atoms.
The potential has a cut-off radius $2.5 \sigma$ where $\sigma$ is the diameter of the atoms.
The parameters $\epsilon$ and $\sigma$ are taken to be the same for both phases.
The parameter $c_{\alpha \beta}=1$ if $\alpha=\beta$; $c_{\alpha \beta}=c_{FS}$
  otherwise controls the wetting interaction between the fluid and the solid nanoparticle.  In addition to the Lennard-Jones
 interactions, atoms inside the  particles are connected to their neighbors with
  FENE springs~\cite{vladkov2006} $V(r)=-0.5 k R_0^2 \ln \left( 1-(r/R_0)^2 \right)$ with $k=30 \epsilon/\sigma^2$ and
  $R_0=1.5 \sigma$.  This nearest neighbor bonding allows one to heat up the nanoparticle
  to rather high temperatures without observing the melting or fragmentation phenomena mentioned above..
  In the following, we will concentrate on results obtained for $c_{FS}=1$, which will be shown in section \ref{cs} to be an appropriate value
  for a mapping between the LJ and gold-octane systems. The dependence on $c_{FS}$ is discussed in ref. \cite{merabia2009}.
  All results in this section are given in standard Lennard-Jones units, $\epsilon/k_B$, $\sigma$ and $\tau= \sqrt{m\sigma^2/\epsilon}$
  for temperature, length and time, respectively.
 Here $m$ is the mass of the fluid atoms.

We integrate the equations of motion using a velocity Verlet algorithm with a time step
$dt=0.005 \tau$. All the systems considered have been first equilibrated at a constant temperature $T_0=0.75$ under the
constant pressure $P_0 =0.015$ (using a Nose/Hoover temperature thermostat and pressure
 barostat). The temperature $T_0$ is below the boiling temperature,
that we found to be $T_b \simeq 0.8$, using independent simulations of a liquid/vapor interface,
under the pressure $P_0$ we are working at. After $100000$ time steps of equilibration, the nanoparticle is heated up
 at different temperatures $T_p>T_b$ by rescaling the velocities of the solid particles
 at each time step, while the whole system is kept at the constant pressure $P_0$
 using a NPH barostat. The fluid beyond a distance $10 \sigma$ from the particle
 surface is thermostatted at $T_0=0.75$, again using velocity rescaling.
Temperature, density and pressure fields have been obtained by averaging the corresponding quantities during
 $10000$ time steps in nanoparticle centered spherical shells of width $\simeq 0.15 \sigma$, after a steady state is reached.
  Finally, we calculate the heat flux density
  flowing through the solid particle, by measuring the power supply needed to keep the nanoparticle at the
  target temperature $T_p$.
\begin{figure}[htp]
\begin{center}
\includegraphics[width=6cm]{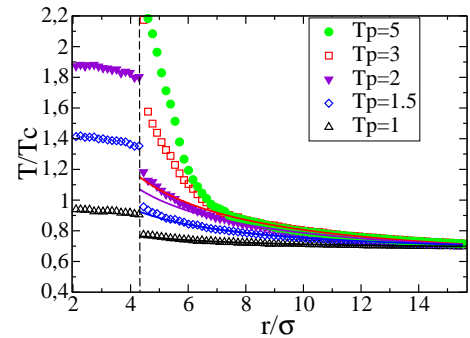}
\caption{\label{temperatureprofile} (Color online) Steady state temperature field
across the liquid/nanoparticle interface, obtained with the Lennard-Jones model (adapted from the results of ref.
\cite{merabia2009}.
The temperature has been normalised by the critical temperature $T_c$ of the  Lennard-Jones fluid.
The position of the nanoparticle surface is indicated by dashed lines, $r$ measuring the distance
to the center of the nanoparticle.
Solid curves correspond to fits by the continuum theory result
$T(r)=A/r+B$. Note that for $T_p=1$ and $T_p=1.5$, the solid curves are almost
indistinguishable from the simulation data. For the highest temperatures, the temperature field inside the
particle is not shown, in order to limit the amplitude of the scale of the vertical axis.}
\end{center}
\end{figure}
Figure~\ref{temperatureprofile} displays steady state
temperature profiles close to the nanoparticle surface, for different
temperatures $T_p$ of the nanoparticle. For low $T_p$, the temperature field
in the liquid is practically indistinguishable from the form $A/r+B$ (eq.~\ref{temp_spherical}).
The general behavior is strikingly similar to the one obtained for the gold-octane system.
Inside the solid, the temperature is not uniform but slightly curved downwards,
due to the finite conductivity of the nanoparticle.
As in the gold/octane simulation, the temperature is discontinuous at the interface. For a given value
of the nanoparticle temperature $T_p$, the relative
temperature jump $\Delta T$ is quite comparable to the gold-octane results, thus suggesting that
the corresponding interfacial conductance is depends
on the details of the nanoparticle/fluid interaction mostly trough macroscopic parameters such as  the wettability \cite{xue2003,barrat2003,ge2006}..
Similar indications can be found in Ref. \cite{shenogina2009}, where it was shown that the interfacial conductance of surfactant-water interfaces is
directly proportional to the work of adhesion.
For instance, for the nanoparticle heated up at $T_p=1.5$,
we have measured a temperature jump $\Delta T/T_c \simeq 0.4$ which is quite comparable to the value reported in
fig.~\ref{conductance_gold} for the gold nanoparticle for the $400$ nW heating power. Similarly, the value of the conductance obtained
 $G=j_Q/\Delta T \simeq 0.6$ is consistent with the gold/octane results displayed in fig.~\ref{conductance_gold}, if we assume that a value of $G=1$ in our LJ units corresponds to an interfacial conductance on the order of $100$ MW/K/m$^2$.
All these results suggest that the interfacial conductance is a quantity which does not strongly depend
on the details of the nanoparticle/fluid interaction but only on generic properties of an interface as {\em e.g} the wettability. This correspondance between
the simplified  Lennard-Jones system and the more realistic description of gold in octane will be explored further in the next section.

We also  mention briefly that very high flux situations  can be explored here
thanks to the "covalent" bonding introduced between the atoms of the solid particle.
 Upon increasing the temperature of the nanoparticle,
deviations from the $1/r$ behavior are clearly seen in fig.~\ref{temperatureprofile},
when the local temperature exceeds the critical temperature $T_c$. Interestingly,
the temperature profile steepens close to the nanoparticle surface,
corresponding to a decrease of the local effective conductivity.. The density profile
\cite{merabia2009} differs
somewhat from its octane or water counterparts, with  density oscillations in the case
of the monoatomic LJ fluid. These oscillations are smeared out for
a molecular liquid. The most important point  is that, even far above the critical point, we do not observe
a steep decrease of the liquid density in the vicinity of the
particle, but rather the appearance of a dilute liquid layer, with a density decreasing when the nanoparticle gets hotter.
 Note however that the density within this layer is still one order of magnitude larger than the vapor density at coexistence.
 Thus, even at temperatures several times $T_c$,
 boiling of the surrounding fluid is not observed.

\section{Corresponding states analysis}
\label{cs}
Obviously, the qualitative similarities between the observations in the previous sections points to a rather generic scenario.
In order to allow a more quantitative comparison, a mapping between the different
systems is necessary. The mapping we investigate will be based on the physical properties that dominate the
problem under consideration, namely  interfacial effects, liquid vapor coexistence and
heat transfer in the liquid phase.

We start by a discussion of the gold-octane case, which in view of the rather similar interactions can be expected to be easily mapped
onto a Lennard-Jones system.
The first step is to tune the interaction coefficient between fluid and solid, $c_{FS}$, to a value that is best suited to reproduce the
properties of the gold octane interface. Generally speaking, the wetting properties of an interface can be related to the
interaction potential $u_{ij}(r)$ through the interfacial work~\cite{widom1982} $H_{ij}=-\frac{{\rho}_i {\rho}_j} {4}\int_{r_0}^{+\infty} r u_{ij}(r) d\mathbf{r}$ where $r_0$ is a minimal radius of approach between two molecules, $\rho_i$ and $\rho_j$ denoting the number densities of the interacting media. The wetting
 properties (equilibrium contact angle or spreading parameter) will be determined by the ratio $r= H_{FS}/H_{FF}$. In the gold octane system, this ratio is
 : $  \frac{\rho_{Au}}{\rho_{CH2}}\frac{\epsilon_{CH2/Au}}{\epsilon_{CH2/CH2}}
\frac{\sigma_{CH2/Au}^6}{\sigma_{CH2/CH2}^6}$.
In the latter expression, $\rho_{Au}=0.1$ mol.cm$^{-3}$ and $\rho_{CH_2}=0.05$ mol.cm$^{-3}$ are the gold and fluid  number densities,
 $\epsilon_{CH_2/Au}$=0.429 kcal.mol$^{-1}$; $\epsilon_{CH_2/CH_2}$=0.143 kcal.mol$^{-1}$ are the octane united atoms/gold and octane/fluid
interaction energies \cite{xia1992}, while $\sigma_{CH_2/Au}$=0.328 nm; $\sigma_{CH_2/CH_2}$=0.3923 nm are the radii of the corresponding interaction. Hence we find $r\simeq 2 $. For the monoatomic LJ model, $r= c_{FS} \rho_S/\rho_F$ where the nanoparticle and liquid densities are, respectively, $\rho_S \simeq 1.46 \sigma^{-3}$ and $\rho_L \simeq 0.76 \sigma^{-3}$. To match the value of $r$, a value of parameter $c_{FS}\simeq 1$, as used in section \ref{lj}, is therefore appropriate.
.

Now, we discuss the value of the unit of thermal flux used in the generic LJ model. To this end, we shall first determine the units of length, time and energy $\sigma$, $\tau$, and $T$ corresponding to the generic model.
Throughout, we denote by stars quantities expressed in LJ units where the units of length $\sigma$, time $\tau$ and energy $\epsilon$ are all set equal to $1$. Alternatively, we can determine the values of $\sigma$, $\tau$ and $\epsilon$ by matching the thermal properties of the generic LJ model to the ones of the gold/octane model. By matching the critical temperature of the LJ model, $T_c^{\star} =1.08$ to the octane critical temperature $T_c=569$ K, we obtain the unit of energy, $\epsilon/k_B=527$ K. The values of the units of length and time are obtained by matching the values of the thermal conductivity $\lambda$ and thermal diffusivity $D_{th}$ of octane~
$\sigma^3 = \frac{D_{th} k_B}{\lambda} \frac{{\lambda}^{\star} {{\sigma}^{\star}}^3 } {D_{th}^{\star} k_B^{\star}}$ and
$\tau = \frac{k_B}{\lambda \sigma} \frac{\lambda^{\star} \sigma^{\star} \tau^{\star}}{k_B^{\star}}$. The thermal conductivity of the monoatomic LJ fluid $\lambda^{\star}=0.36$ was measured using stationary heat transfer simulations, while we have used the value $D_{th}^{\star}=1$ for the thermal diffusivity reported in \cite{palmer1994}. Using the values of the thermal conductivity $\lambda = 0.1$ W.m$^{-1}$.K$^{-1}$ and the thermal diffusivity $D_{th}=6.4 \quad 10^{-8}$ m$^{2}$.s$^{-1}$ at $400$ K~\cite{watanabe2002}, we obtain $\sigma =0.32$ nm and $\tau =1.6$ ps. The power of 1 in Lennard Jones units corresponds then to  $\frac{k_B T_c}{\tau}=5$ nW, while a boundary conductance $G^*=1$ is equivalent to a real $G=k_B/\tau\sigma^2 = 88$MW/K/m$^2$.
\begin{figure}[htb]
\begin{center}
\includegraphics[width=6cm]{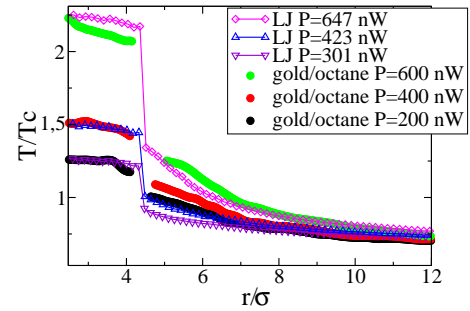}
\caption{\label{comparison_temp_profile} (Color online) Reduced temperature profiles across the gold/octane interface (filled symbols) compared to the profiles obtained with the LJ model (open symbols). $T_c$ is the critical temperature of octane and of the LJ fluid respectively. For the gold/octane system, $\sigma=0.32$ nm as discussed in the text. Here the gold/octane system was thermostatted at $T=380$ K.}
\end{center}
\end{figure}
To illustrate the relevance of the mapping discussed, we have compared in fig.~\ref{comparison_temp_profile} the temperature profiles
corresponding to the gold/octane interface and to the LJ model, in terms of the reduced temperature $T/T_c$. The distances have been rescaled here by the value of $\sigma=0.32$ nm discussed before. For the sake of the comparison,
the gold/octane systems have been thermostatted at a higher temperature
$T=380$ K than before, since at the reduced temperature $T=300/569$, the LJ
fluid may cristallise. The agreement between the atomically realistic model
and the monoatomic LJ model is fairly good. For the lower heating strengths
considered, the LJ fluid develops almost the same temperature profile away
from the nanoparticle. In particular, the slopes of the temperature profile
(flux) at the solid interfaces are quite comparable. Note however, that the
temperature jump $\Delta T$ at the interface is larger in the case of the LJ
model, a fact which can be attributed to the slightly smaller values of the
interfacial conductances of the monoatomic model compared to the more
realistic system. For higher heating powers, the LJ temperature lies slightly
above the octane curve. However, the fluxes at the solid interfaces are again
quite comparable and the interfacial temperature jumps $\Delta T$ compare
well, probably due to the rapid decrease of the gold/octane conductance with
the supplied power.

We have carried out a similar analysis for the gold/water system. The values
retained are $\lambda = 0.58$ W/m/K  and $D_{\rm th}=1.3$ $10^{-7}$ m$^²$/s for
the conductivity and thermal diffusivity at $T=450$ K, and $T_c=650$ K for the
critical temperature. This yields $\sigma=0.22$ nm, $\tau=0.384$ ps and  $\epsilon/k_B=602$ K. for the
unit of length, time and energy, respectively. The values of the length and time units are smaller than their
gold/octane counterpart, due to the larger thermal conductivity of water.
As a consequence, the units of power $P=21.6$ nW and interfacial conductance
$G=537$ MW/m$^{2}$/K are larger than in the previous case. Because of
the complexity of the interatomic potentials involved in the water/gold
system, we have used a different approach than in the gold/octane case  to adjust the solid-liquid interaction parameter for the
"corresponding" Lennard-Jones system. The reduced temperatures $T/T_c$ of the nanoparticles
where chosen to be identical, and the
parameter $c$ was chosen such that the power input is similar in both cases. This essentially amounts to
matching the interfacial conductances of the two systems.
  The resulting value $c=0.78$ leads to an interfacial work
ratio $r=1.56$. This is slightly different from the value  $r=1.9$  based on
the gold/water interfacial work $\pi \rho_{Au} \rho_O \epsilon_{O/Au}
\sigma_{O/Au}^4 /3 \simeq 0.13$ kcal/mol/A$^2$ and the water surface tension
$\gamma \simeq 40$ mJ/m$^2$ at $450$ K\cite{vega}.  This difference is probably due to the more complex structure
of the gold-water interface, which implies that a simple matching based on identifying the $r$ parameters is not appropriate.

\begin{figure}[htb]
\begin{center}
\includegraphics[width=6cm]{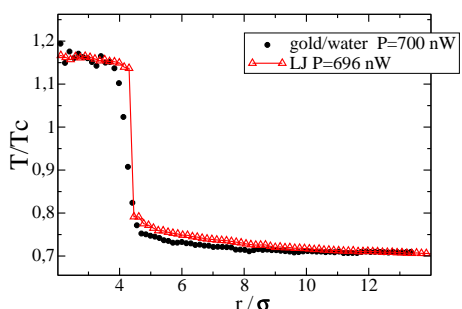}
\caption{\label{comparison_temp_profile_water} (Color online) Reduced
  temperature profiles across the gold/water interface (filled symbols)
  compared to the profiles obtained with the LJ model (open symbols). Here
  $T_c=650$ K and $\sigma=0.22$ nm for water.}
\end{center}
\end{figure}

When the matching is performed on the interfacial conductance,
the agreement between the two temperature profiles,  shown
in fig.\ref{comparison_temp_profile_water}, is very good. . That this is the case may seem
obvious, as the main macroscopic parameters have been matched. However, the fact that the agreement
is obtained  down to subnanometer scales, in a situation where heat fluxes are extremely strong, and can be transferred to
different values of the power input, is far from trivial.
In conclusion, the simple monoatomic LJ  model, with an appropriate choice of the parameters, can be used
le to reproduce quantitatively the features of heat transfer around a nanoparticle obtained for systems with more complex interactions.
It constitutes a simple and efficient tool to explore heat transfer at the nanoscale, in th espirit of a coarse graining approach of such mesoscale
phenomena.

\section{Conclusions}

We have explored the phenomenon of heat transfer  in the vicinity of strongly heated nanoparticles,
using molecular dynamics simulations of atomically realistic models or of more coarse grained Lennard-Jones monoatomic fluids.
The comparison between the two approaches shows that, provided the mapping is carried out using the physically relevant properties,
they are quantitatively equivalent. The simulations reveal that the fluid in the vicinity of the nanoparticles can sustain very
high heat fluxes and large temperature differences without undergoing the type of drying instability that is observed on flat surfaces, and that
temperatures much above the critical temperatures can be reached without observing phase coexistence
.  In the case of gold in octane,
 high heat fluxes and temperatures can  result in a partial desintegration of the nanoparticle, while
 in the gold in water case they  only result in melting of the particle for comparable heating powers..

The phenomena that are involved in such experiments are quite complex, with a combination of
phase transition, interfacial phenomena and transport phenomena,
all taking place on the nanometer scale. It is therefore not trivial, that a simple coarse graining based on a matching of
interface and thermal properties results in a correct  description of the phenomenon. This implies that
coarse grained methods should be appropriate for describing similar phenomena in nanostructures of larger dimensions
(e.g. aggregates of nanoparticles, or nanostructured surfaces). The level of coarse graining and the gain in efficiency
provided by the monoatomic LJ fluid could be even improved by extending to appropriately adapted versions of other
 coarse graining approaches such as free energy models \cite{joly2008} or
appropriate versions of dissipative particle dynamics \cite{dpd}.





\begin{acknowledgments}
We acknowledge financial support from ANR project Opthermal, and many fruitful
interactions with  L.J. Lewis. Part of the simulations wer realized using the LAMMPS package \cite{lammps}
\end{acknowledgments}





\end{article}








\end{document}